\documentclass[12pt]{emulateapj}

\usepackage[dvips]{color}

\shorttitle{New class of VHE emitter: mini-shells in AGNs}
\shortauthors{Kino et al.}

\begin{document}

\title{New class of Very High Energy $\gamma$-ray emitter: 
radio-dark mini-shells surrounding AGN jets}
\author{\thanks{Last update:7 th Sep, 2012}
Motoki Kino\altaffilmark{1}, 
Hirotaka Ito\altaffilmark{2}, 
Nozomu Kawakatu\altaffilmark{3}, and
Monica Orienti\altaffilmark{4, 5}
}

\altaffiltext{1}{ISAS/JAXA, 3-1-1 Yoshinodai, 229-8510 Sagamihara, Japan} 
\email{kino@vsop.isas.jaxa.jp}
\altaffiltext{2}{Yukawa Institute for Theoretical Physics, 
Kyoto University,
Oiwake-cho Kitashirakawa Sakyo-ku, Kyoto, 606-8502, Japan}
\altaffiltext{3}{Graduate School of Pure and Applied Sciences,
University of Tsukuba, 1-1-1 Tennodai, Tsukuba 305-8571, Japan}
\altaffiltext{4}{Dipartimento di Astronomia, Universita di Bologna,
via Ranzani 1, I-40127, Bologna Italy}
\altaffiltext{5}{
INAF - Istituto di Radioastronomia, 
via Gobetti 101, 40129 Bologna, Italy}



\begin{abstract}

We explore non-thermal emission from  
a shocked interstellar medium, which is
identified as an expanding shell, 
driven by a relativistic jet in active galactic nuclei (AGNs). 
In this work, we particularly
focus on parsec-scale size mini-shells surrounding
mini radio lobes.
From radio to X-ray band,
the mini radio lobe emission dominates 
the faint emission from the mini-shell.
On the other hand,
we find that inverse-Compton (IC) emission from the shell
can overwhelm the associated lobe emission 
at very high energy (VHE; $E>100~{\rm GeV}$) 
$\gamma$-ray range, because energy densities of 
synchrotron photons from the lobe and/or
soft photons from the AGN nucleus are large
and IC scattering effectively works.
The predicted IC emission from nearby mini-shells 
can be detected with the {\it Cherencov Telescope Array} (CTA)
and they are potentially a new class of VHE $\gamma$-ray 
emitters.

\end{abstract}


\section{Introduction}
\label{sec:intro}

Radio-loud active galactic nuclei (AGNs)
are among the most powerful objects in the Universe.
According to the standard picture of 
jets in AGNs, the jets are enveloped
in a cocoon consisting of shocked jet material
and the cocoon is surrounded by
shocked interstellar medium.
The shocked ambient region 
(hereafter we refer to as the shell) 
is identical to the forward shocked region and it is 
a fundamental ingredient in 
the whole AGN jet system.
In spite of this, physical properties of shells have
not been well studied since they are not bright and 
remain undetected except for 
thermal X-ray detections of shells surrounding bubbles 
in Centaurus A (Croston et al. 2009),
NGC3801 (Croston et al. 2007),
and in the Galactic center (Su et al. 2010 and reference therein).
In radio band, shells are dim (Carilli et al. 1988) 
and their overall emissions are overwhelmed by the 
radio bubbles (radio lobes) identical to 
a portion of the cocoon.

Recently, we indicate 
a possibility of VHE $\gamma$-ray emissions 
from AGN shells
in Ito et al. (2011) (hereafter I11).
In I11, 
the non-thermal emission of a shell is mainly produced 
by Inverse-Compton (IC) mechanism
for compact sources (less than $\sim10$~kpc), 
while synchrotron radiation is more important
for larger shells.
Physical properties of nearby shells can be proved
by the detection of IC emissions using 
modern Cherenkov telescopes.
The examined shell size in I11 was, however, 
limited to $1-100$~kpc.
In this paper, we complement the work of I11
by exploring smaller shells. 
I11 indicates that smaller shells may be 
a new class of VHE $\gamma$-ray emitter in the universe
and in the present work we quantitatively examine the 
theoretically predicted photon spectra from
the mini-shells.
To clarify the detection  feasibility
by 
the next generation instrument 
{\it Cherencov Telescope Array} (CTA), 
we further examine the non-thermal emission on
$10$~pc scale, which corresponds to the smallest scale of  
radio bubbles ever observed, properly
taking into account the $\gamma\gamma$ absorption
by extragalactic background light (EBL).
The CTA will consist of two arrays 
of Cherenkov telescopes, which aim to: 
(a) increase sensitivity by one order of 
magnitude for deep observations around 1~TeV, 
(b) boost significantly the detection area 
and hence detection rates, 
(c) increase the angular resolution and hence
the ability to resolve the morphology of extended sources, 
(d) provide uniform energy coverage for photons from some 
tens of GeV to beyond 100~TeV, and 
(e) enhance the sky survey capability, 
monitoring capability and flexibility of operation 
(e.g., Actis et al. 2011; Funk et al. 2012).

In this work,
we particularly focus on the nearby mini radio lobes 
of CORALZs (COmpact RAdio sources at Low-Redshift) 
which are located at the redshift of $0.005\le z\le 0.16$
(Snellen et al. 2004; de Vries  et al. 2009) 
and less luminous mini lobes which would be detected 
by deep sensitivity observations
in the future 
by using the next generation radio telescope
{\it Square Kilometer Array} (SKA)
with its collecting area distributed over
a large geographical area 
(Lazio 2011 for details of its specification).




\section{Model}\label{model}

\begin{figure}
\includegraphics[width=8cm]{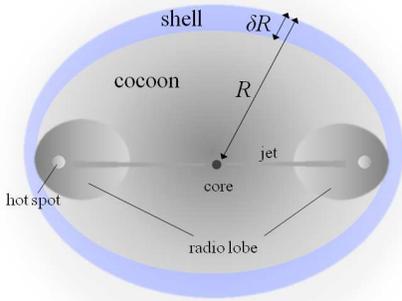}
\caption{A cartoon of
relativistic jets and ambient matter interaction. 
The kinetic energy of the jets
is dissipated via shocks at the hot spots 
and deposited into the cocoon 
with its radius $R$ and the shell with its width $\delta R$. 
The cocoon is inflated by its internal energy.}
\label{fig:cartoon}
\end{figure}

First, we briefly review the model 
following our previous work I11.
For simplicity, 
we neglect the elongation of cocoon and shell 
and we adopt the expanding spherical
bubble model 
(see Fig. \ref{fig:cartoon}).
The shell width at the bubble radius $R(t)$ 
at the time ($t$) is denoted as $\delta R(t)$.
The mass density of the ambient matter 
at $R(t)$ is defined as 
$\rho_{\rm a}(R(t))=\rho_{0}(R(t)/R_{0})^{-\alpha}$
($0 \leq \alpha \leq 2$)
where $R_{0}$ is the reference radius, 
$\delta R(t)$ satisfies the relation 
$\delta R(t)=
(\hat{\gamma_{\rm a}}-1)R(t)/[(\hat{\gamma_{\rm a}}+1)(3-\alpha)]$
where $\hat{\gamma_{\rm a}}$ is the specific heat ratio
of the ambient matter (e.g., I11).
 The thin shell condition $\delta R(t)\ll R(t)$ holds
 when the expansion velocity has a high Mach number.
We further assume that the kinetic 
power of jet, $L_{\rm j}$, is constant in time.     
The jet kinetic energy is dissipated and depositted 
as the internal energy of the cocoon and shell. 
The internal energy drives the cocoon expansion. 
Note that specifying
a fraction of electrons/protons in the jet
is irrelevant to the expansion dynamics.
Solving the set of equations for 
the momentum and energy equations 
for the expanding bubble,
$R(t)$ can be expressed as
\begin{eqnarray}
\label{R}
R(t)=C~~
            {R_0}^{\alpha/(\alpha-5)}~
             \left(\frac{ L_{\rm j}}{\rho_0} \right)^{1/(5-\alpha)}
          t^{3/(5-\alpha)} ,
\end{eqnarray}
where the coefficient $C$ is given by
$C = 
\left[ \frac{(3-\alpha)(5-\alpha)^3(\hat{\gamma_{\rm c}}-1)}
{4\pi \{ 2\alpha^2 + (1-18\hat{\gamma_{\rm c}})\alpha + 63\hat{\gamma_{\rm
c}}-28 \}  }
 \right]^{1/(5-\alpha)}$
where  $\hat{\gamma_{\rm c}}$ is 
the adiabatic index of the cocoon.
This is identical to the well known  stellar wind model
(e.g., Castor et al. 1975;  Ostriker and McKee 1988).
%
For convenience, we define
the total internal energy deposited in the shell and cocoon
as follows:
$E_{\rm shell} = f_{\rm shell} L_{\rm j} t$, and
$E_{\rm cocoon} = f_{\rm cocoon}L_{\rm j} t$
where the factors $f_{\rm shell}$ and $f_{\rm cocoon}$ 
can be given  by 
$ f_{\rm shell} 
= \frac{18 ({\hat{\gamma}_{\rm c}} - 1) (5-\alpha)}
           {({\hat{\gamma}_{\rm a}} + 1)^2
             [2\alpha^2 + (1-18{\hat{\gamma}_{\rm c}}) \alpha +
              63 {\hat{\gamma}_{\rm c}} - 28]}$ 
and
 $f_{\rm cocoon} = (5 - \alpha) (7 - 2 \alpha)/
 [2 \alpha^2 + (1 - 18\hat{\gamma}_{\rm c})\alpha + 63 \hat{\gamma}_{\rm c} - 28] $.
Hereafter we assume typical cases of 
${\hat \gamma}_{\rm c} = 4/3$, 
${\hat \gamma}_{\rm a} = 5/3$ and
$\alpha = 0$.
Regarding ambient matter, 
we set
$\rho_{0}=0.1m_{p}~{\rm cm^{-3}}$, and
$R_{0}=1~{\rm kpc}$ which are typical values 
for elliptical galaxies
(Mathews and Brighenti 2003 for review).
The value of magnetic field strength 
in elliptical galaxies is $\sim$ a few ${\rm \mu G}$ 
(Carilli and Taylor 2002 for review).
Therefore, we fix the magnetic field strength
in the shell as $B_{\rm shell} = 10{\mu {\rm G}}$ 
where we assume the shock compression ratio being four.

Second, we show the basic treatment of
photon and electron distributions in shells and lobes.
We solve the following  kinetic equation
describing the electron 
energy distribution $N_{e}(\gamma_e, t)$ 
as follows:  
\begin{eqnarray}
 \label{kinetic}
\frac{\partial N_{e}(\gamma_{e}, t) }{\partial t} =
 \frac{\partial}{\partial \gamma_e}
  [ \dot{\gamma}_{\rm cool}(\gamma_{e},t) N_{e}(\gamma_{e}, t) ]
 + Q_{e}(\gamma_e,t) ,
\end{eqnarray}
where 
$\gamma_e$, $\dot{\gamma}_{\rm cool}(\gamma_{e},t) 
= - d\gamma_e/dt
=
\dot{\gamma}_{\rm ad}+
\dot{\gamma}_{\rm syn}+
\dot{\gamma}_{\rm IC}$,
and 
$Q_{e}(\gamma_{e},t)\propto \gamma_{e}^{-p}$
are the Lorentz factor, 
the cooling rate via adiabatic expansions 
($\dot{\gamma}_{\rm ad}={\dot R}\gamma_{e}/R$)
and radiative losses
and the injection 
rate of non-thermal electrons, respectively.
Regarding IC scattering, we take the
Klein-Nishina cross section into account and
the following seed photons are included: 
(1) UV photons from a standard accretion disk,
(2) IR photons from a dust torus,
(3) synchrotron photons from the radio lobes, 
(4) synchrotron photons from the shell, and
(5) Cosmic Microwave Background (CMB).
As for (5), the energy density of CMB photons 
is sufficiently small in the present case, 
so we neglect it.


Third, we include the effect of absorption 
via $\gamma \gamma\rightarrow e^{\pm}$ interaction.
Very high energy photons
suffer from the  $\gamma \gamma$ absorption
via interaction with various soft photons
(e.g., Coppi and Aharonian 1997).
Here we include the $\gamma \gamma$ absorption
due to both source-intrinsic and EBL photon fields.
The $\gamma\gamma$ absorption opacity
against the intrinsic photons 
($\tau_{\gamma\gamma}$) can be 
calculated by summing up all of the photons 
from 
(1) the shell, 
(2) the radio lobes, 
(3) the dusty torus, and 
(4) the accretion disk
and we multiply 
the $\gamma\gamma$ absorption factor of 
exp$(-\tau_{\gamma\gamma})$ to 
the unabsorbed flux.
For simplicity,
we deal with the $\gamma\gamma$ absorption effect
at the first order and we neglect cascading effect.
With regard to the opacity for
$\gamma\gamma$ 
interaction between EBL and TeV photons, 
we adopt the model of  
Franceschini et al. (2008). 
They use the available information on extragalactic
sources generating diffuse photons in the universe
between far-UV and the sub-millimeter ranges
and 
the opacity-corrected TeV blazars spectra are 
consistent with standard photon generation processes 
which lead to intrinsic photon indices 
steeper than  $\Gamma_{\rm intrinsic}=1.6$
(e.g., Aharonian et al. 2006 and references therein).

The model parameters are as follows.
As for radio lobes, the fractions  of non-thermal 
electron energy and magnetic energy are defined as
$\epsilon_{\rm e,lobe}$ and
$\epsilon_{\rm B,lobe}$, respectively.
The spectral index,
the gyro factor,
the minimum and 
maximum Lorentz factors of non-thermal electrons
are denoted as 
$p_{\rm lobe}$,
$\xi_{\rm lobe}$, 
$\gamma_{\rm lobe,min}$, 
and $\gamma_{\rm lobe,max}$, respectively.
As for a shell, 
the fraction  of non-themal 
electron energy to the total internal energy 
is defined as $\epsilon_{\rm e,shell}$.
The spectral index,
the gyro factor,
the minimum and 
maximum Lorentz factors of non-thermal electrons
are expressed as 
$p_{\rm shell}$,
$\xi_{\rm shell}$, 
$\gamma_{\rm shell,min}$, 
and $\gamma_{\rm shell,max}$, respectively.
Using the electron gyro factor 
$\xi_{i}$ ($i=$lobe or shell), 
the electron acceleration rate with its energy
$\gamma m_{e}c^{2}$ is given by
$d\gamma/dt =3eB{\dot R(t)}^{2}/20\xi_{i} m_{e}c^{3}$.
Hereafter we assume
$\gamma_{\rm lobe,min}=\gamma_{\rm shell,min}=1$ .
The observed range of the size $R(t)\equiv LS/2$ 
can be given by $5~{\rm pc}\le R(t) \le 175~{\rm pc}$
which correspond
to LS/2 of the most compact and most extended sources in the CORALZ
sample, i.e. J103719+433515 and J160246+524358 (de Vries et al. 2009).
In this paper, we demonstrate one typical
case of $R(t)=5~{\rm  pc}$ with $z=0.08$.
Observational constraints on $L_{\rm j}$ and 
$UV$ and $IR$ luminosities at the nucleus  
($L_{\rm UV}$ and  $L_{\rm IR}$)
are discussed in the next section.


\section{Observational Constraints}

Here we show key observational constraints
which restrict the model parameters.

\subsection{Nucleus luminosities: $L_{\rm UV}$ and $L_{\rm IR}$}

The luminosities $L_{\rm UV}$ and $L_{\rm IR}$ are
important as the seed photons for IC scattering. 
In young radio sources, it is 
difficult to estimate them  directly
because they are heavily absorbed
by the dusty torus 
(e.g., Kawakatu et al. 2009; 
Ostorero et al. 2010).
Prior works
(e.g., 
Snellen et al. 1999;
Vink et al. 2006;
Orienti et al. 2010)
indicate that faint compact radio sources
are generally characterized by a low level of ionization.
Mack et al. (2009) observed the CORALZ sample at 250~GHz 
with the IRAM 30-meter telescope to search for dust presence
and they indeed
found dust emission in a significant fraction of the sample.

Here we estimate $L_{\rm UV}$ and $L_{\rm IR}$
by means of 
observations of other compact radio sources.
Kunert-Bajraszewska and Labiano (2010; hereafter KL10) 
show that low-luminosity
compact radio sources' [OIII]$\lambda$5007 luminosity 
$L_{\rm [OIII]}\sim 10^{40}-10^{43}~{\rm erg~s^{-1}}$.
Radio luminosity of KL10 sample is comparable with CORALZ one.
Hence, in the case of CORALZ sample, we derive 
the $L_{\rm [OIII]}$
by assuming the relation of $L_{\rm radio}$
and $L_{\rm [OIII]}$ obtained 
for low redshift radio galaxies
(e.g., Buttiglione et al. 2011; Son et al. 2012),
and we obtain
$L_{\rm [OIII]}\sim 10^{40}-10^{43}~{\rm erg~s^{-1}}$.
Since disk's UV emission
is likely a main source for the ionization 
of clouds in narrow line regions,
it is reasonable to suppose that 
$L_{\rm UV} > L_{\rm [OIII]}$
although it is not very clear 
what the re-emitting fraction is.
Here we examine the case of 
$L_{\rm UV}=6\times10^{42}, 6\times10^{43}, 
6\times10^{44}~{\rm erg~s^{-1}}$.
Regarding the torus luminosity,
Calderone et al. (2012) 
explore the fraction of torus 
re-emission of absorbed 
accretion disc radiation for 
about 4000 radio-quiet AGNs and 
they found that the 
torus reprocesses $\sim 1/3-1/2$ of the accretion disk
luminosity. Based on their work,
we assume $L_{\rm IR}=L_{\rm UV}/2$.
The IR and UV photon energy densities
at $R(t)$ are given by 
$U_{\rm IR}(t)=L_{\rm IR}/4\pi c R(t)^{2}$ and
$U_{\rm UV}(t)=L_{\rm UV}/4\pi c R(t)^{2}$, respectively.

We note that, in a few objects, a fraction of 
IR seed photons may arise from a dense cocoon of 
dust likely deposited by a merger event 
(e.g., Holt et al. 2009).
We do not treat this case merely for simplicity.

\subsection{Radio lobe fluxes}

{\it Optical}.
The CORALZ sample have been 
identified with bright galaxies
in Automated Plate Measuring machine
catalogue of the first Palomar Observatory Sky
Survey 
and some further identifications 
in the 6th Sloan Digital Sky Survey are added and 
the CORALZ sample increases in number 
(de Vries et al. 2009 and references therein).
Since mini lobes are smaller than 
their host galaxies, it is hard to measure 
the pure emission of mini lobes.

{\it X-ray}.
No X-ray observations of the
CORALZ sample have been performed so far. 
Although the emission mechanism has not yet been 
confirmed (i.e., accretion disk or radio lobes),
X-ray fluxes of 
gigahertz-peaked spectrum (GPS)
and compact steep-spectrum (CSS) sources
have been measured by several authors recently
(e.g., Tengstrand et al. 2009).
Here we regard the measured X-ray fluxes in GPS galaxies
$0.7\times 10^{-14}~{\rm erg~cm^{-2}~s^{-1}}\le 
\nu F_{\nu}\le 
5.6\times 10^{-13}~{\rm erg~cm^{-2}~s^{-1}}$
(Guainazzi et al. 2006)
as a rough upper limit of X-ray emissions from 
mini-radio-lobes and mini-shells.

{\it GeV $\gamma$-ray}.
Prior to the launch of {\it Fermi}, mini-lobes were
predicted to emerge as a new population of
$\gamma$-ray emitter
(Stawarz et al. 2008; Kino et al. 2007, 2009).
With exposure time
currently accumulated by {\it Fermi}/LAT, however,
the majority of mini-lobes remain un-detected
except for NGC~1275 
with $\nu F_{\nu}\sim 4\times 10^{-11}~{\rm erg~cm^{-2}~s^{-1}}$
(Abdo et al. 2009;
Nagai et al. 2010; 
Suzuki et al. 2012;
Aleksi{\'c} et al. 2012) and
possible detection of 4C+55.17 (McConville et al. 2011).
We consider
these detections as the upper limit of shell flux.
Since the measured GeV-$\gamma$
flux of 4C+55.17 $\sim 1\times 10^{-11}~{\rm erg~cm^{-2}~s^{-1}}$
is less luminous than NGC~1275,
we set it as the upper limit.

\subsection{SSA turnover frequencies of radio lobes}

Synchrotron self-absorption (SSA) 
turnover frequencies of CORALZs lobes $\nu_{\rm ssa,lobe}$
are typically less than a few hundred MHz 
(de Vries et al. 2009).
It should be emphasized that
we can constrain  $L_{\rm j}\epsilon_{\rm B,lobe}$ 
from the $\nu_{\rm ssa,lobe}$. 
It is well known that 
the SSA turnover frequency is given by
$\nu_{\rm ssa,lobe}\propto 
B_{\rm lobe}^{1/5}S_{\nu,\rm lobe}^{2/5} R_{\rm lobe}^{-4/5}$.
Since $\nu_{\rm ssa, lobe}$, $S_{\nu,\rm lobe}$, and $R_{\rm lobe}$ 
are
observationally determined, 
$B_{\rm lobe}
\propto (L_{\rm j}\epsilon_{\rm B,lobe})^{1/2}$ is the
only parameter
where
$L_{\rm j}\epsilon_{\rm B,lobe}
\equiv L_{\rm poy}
=4\pi R_{\rm lobe}^{2}c/3\times (B_{\rm lobe}^{2}/8\pi)
=R_{\rm lobe}^{2}c B_{\rm lobe}^{2}/6$
where $L_{\rm poy}$ is the Poynting power
of the jet.
Therefore, we stress that
$L_{\rm j}\epsilon_{\rm B,lobe}$ 
is not a free parameter but a well
constrained quantity.
With typical parameters of mini-lobes,
a large jet power such as
$L_{\rm j}\sim 6 \times 
10^{47}(R_{\rm lobe}/2~{\rm pc})^{2}
(\epsilon_{\rm B,lobe}/10^{-2})^{-1}~{\rm erg~s^{-1}}$
is required.
The requirement of large  $L_{\rm j}\epsilon_{\rm B,lobe}$
for SSA model has been already 
pointed out by several authors
(Fanti et al. 1995; Stawarz et al. 2008).

\section{Results}

The mini-shell parameters can be well 
constrained, since
the forward shocks considered here 
resembles those in the well-studied 
supernovae remnants (SNRs).
Following prior work constraining on 
SNRs parameters
(e.g., Koyama et al. 1995; Ellison et al. 2001),
here we set $\epsilon_{\rm e,shell}=0.05$,
$p_{\rm shell}=2$, and
$\xi_{\rm e,shell}=10$.
Within the observational constraint shown 
in 3.1, we demonstrate the case with 
$L_{\rm j}=5\times 10^{46}~{\rm erg s^{-1}}$.
Shell and radio lobes' kinetic luminosity are 
governed by  
$L_{\rm j}\epsilon_{\rm e,shell}$ and
$L_{\rm j}\epsilon_{\rm e,lobe}$.
Therefore, smaller $L_{\rm j}$ and larger
$\epsilon_{\rm e,shell}$  and $\epsilon_{\rm e,lobe}$
lead to much the same results.

The mini lobe parameters are determined 
on the basis of the observational constraints 
discussed in \S 3.
The radio peak flux densities of CORALZ are about
$\sim 100-500$~mJy at the peak frequency 
$\sim 1$~GHz (de Vries et al. 2009)
and they provide constraints $\epsilon_{\rm e,lobe}$. 
So as not to violate this constraint,
here we set $\epsilon_{\rm e,lobe}=0.01$. 
We assume $p_{\rm lobe}=2.2$ which is
suggested for relativistic shocks 
(e.g., Bednarz \& Ostrowski 1997). 
A large gyro-factor $\xi_{\rm lobe}=10^{7}$ 
is assumed due to the
lack of evidence of synchrotron emission 
from radio lobes at optical band
(e.g., Holt et al. 2007; Fanti et al. 2011).



\begin{figure}
\includegraphics[width=10cm]
{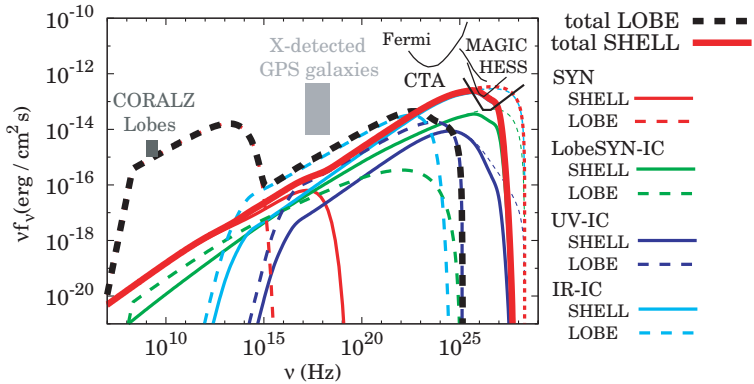}
\caption{
Mini-shell and radio lobe spectra 
(the thick red and black curves, respectively).
Here, we adopt 
$L_{\rm UV}=2L_{\rm IR}=6\times 10^{43}~{\rm erg~s^{-1}}$, and
$\epsilon_{\rm B,lobe}=10^{-3}$.
The IC components of
the lobe synchrotron,
UV from the accretion disk,
IR from the torus, are shown 
in red, green, purple, blue lines, respectively.
Here we plot the radio flux of typical CORALZs
(de Vries et al. 2009) and 
also show X-ray fluxes in GPS galaxies
(Guainazzi et al. 2006) as a reference.
}
\label{fig:typ-spectrum}
\end{figure}
\begin{figure}
\includegraphics[width=8cm]
{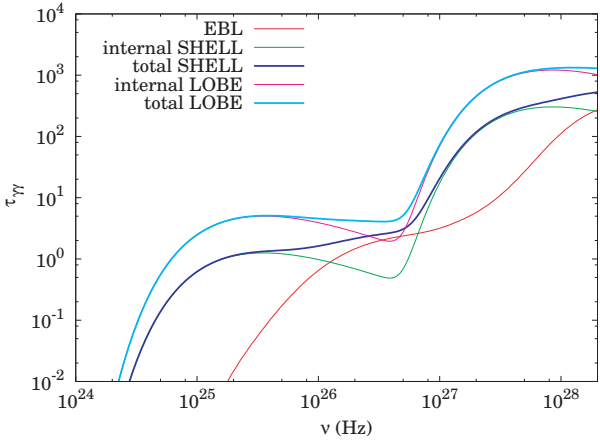}
\caption{ 
The opacity for 
$\gamma \gamma \rightarrow e^{\pm}$
corresponding to the shell spectrum in Fig. \ref{fig:typ-spectrum}.}
\label{fig:opacity}
\end{figure}

In Fig.~\ref{fig:typ-spectrum}, 
we show typical photon spectra
 from mini radio lobes and the mini shell 
with $L_{\rm UV}=2L_{\rm IR}=6\times 10^{43}~{\rm erg~s^{-1}}$, 
and
$\epsilon_{\rm B,lobe}=10^{-3}$.  
We also plot the sensitivities adopted 
from the following web-pages; 
the {\it Fermi}/LAT for 1~yr integration time
 (http://www-glast.stanford.edu/),
 HESS
 (http://www.mpi-hd.mpg.de/hfm/HESS/),
 MAGIC
(http://magic.mppmu.mpg.de/), and 
CTA
(http://www.cta-observatory.org/).
The thick solid and dashed lines colored in red and black 
display the total photon fluxes 
from  the mini shell and mini radio lobe, respectively.
The radio flux density and the SSA turnover frequency of
the mini radio lobes are consistent with CORALZs.
Although the predicted GeV $\gamma$-ray flux appears 
below the {\it Fermi}/LAT sensitivity curve,
the distinctive
double-bump structure in the IC spectrum is found 
reflecting UV and IR emission bumps at the core.
As already shown in I11,
synchrotron emission from the mini-shell 
is very dim. It well explains the 
lack of detection of shells obtained so far.
On the other  hand, bright IC emission is expected
in VHE range and it will be detectable by CTA.
In Fig.~\ref{fig:typ-spectrum}, the mini shell spectrum without
the EBL absorption effect is shown 
in the thin dotted curve.
From this, we see that
the EBL absorption 
is effective above
a few $10^{26}~{\rm Hz}$. 
In Fig.~\ref{fig:opacity}, 
we show the corresponding  optical depth for
$\gamma\gamma$ absorption $\tau_{\gamma\gamma}$ plotted versus
the corresponding high energy photon frequency $\nu$.
At the frequency below 
$\sim 10^{26}~{\rm Hz}$, the number density of
target photons from radio lobes is larger 
than that from the shell,
therefore the photons from the lobe
dominate the absorption opacity.
Since the contribution of the shell's 
synchrotron photons is sufficiently small 
as seed photons of IC and $\gamma\gamma$ absorption, 
we neglect them to save  computational cost.

\begin{figure}
\includegraphics[width=10cm]
{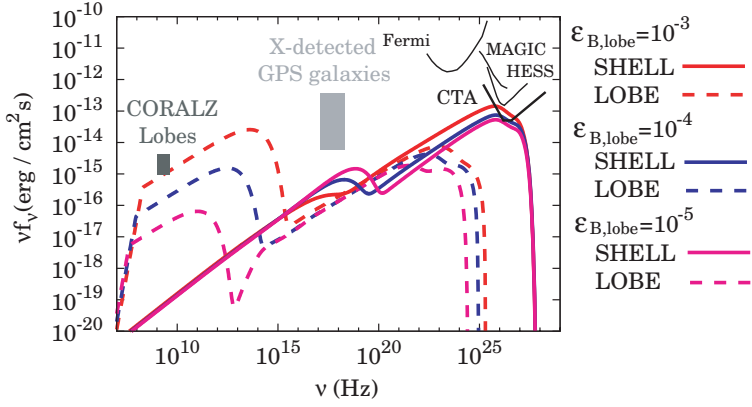}
\caption{
The same mini-shell and radio lobe spectra 
(solid and dashed curves, respectively)
but with
$L_{\rm UV}=2L_{\rm IR}=6\times 10^{42}~{\rm erg~s^{-1}}$, and
$\epsilon_{\rm B,lobe}=10^{-3},10^{-4},10^{-5}$.
The seed photons for IC scatttering in the mini-shell
are dominated by the synchrotron photons of the radio lobes.}
\label{fig:dim-core}
\end{figure}

In Fig.~\ref{fig:dim-core}, we present
predicted spectra with smaller
$L_{\rm UV}=2L_{\rm IR}=6\times 10^{42}~{\rm erg~s^{-1}}$, and
$\epsilon_{\rm B,lobe}=10^{-3}, 10^{-4}, 10^{-5}$.
Here, $\epsilon_{\rm B,lobe}$ has been carefully
chosen not to exceed the observed radio lobe 
luminosities, i.e., 
$\sim (1-5)\times 10^{-15}~{\rm erg~s^{-1}~cm^{-2}}$
at $\sim 1$~GHz.
If $L_{\rm UV}$ becomes smaller, then
the lobe becomes synchrotron-dominated
and the synchrotron luminosity becomes larger
even for the same $\epsilon_{\rm B,lobe}$. 
Hence, the radio flux tends to be overproduced.
We therefore search parameter values 
according to $\epsilon_{\rm B,lobe}\propto L_{\rm UV}$
so as not to overproduce the lobe radio flux.
Overall features of shell spectra are
similar to the case in Fig.~\ref{fig:typ-spectrum}, and 
the mini-shell spectrum
is detectable by CTA.
However the radio lobe IC emission is 
less luminous than the ones in Fig.~\ref{fig:typ-spectrum}
at GeV band simply because of smaller 
$L_{\rm UV}=2L_{\rm IR}$.
Seed photons for IC scattering in the mini-shell
are dominated by the synchrotron photons
from the lobes.
In IC-dominated regime, the peak luminosity 
of IC emission is proportional only to
the $L_{\rm j} \epsilon_{\rm e,shell}$ (see Eq. (24) in I11)
and it only has a weak
dependence to $L_{\rm UV}$ and $L_{\rm IR}$.
The shell luminosity at TeV range becomes slightly 
brighter as the synchrotron emission of the radio
lobes increases.
Therefore, our prediction of emission 
spectra does not strongly depend on
the choice of $L_{\rm UV}$ and $L_{\rm IR}$.

\begin{figure}
\includegraphics[width=10cm]
{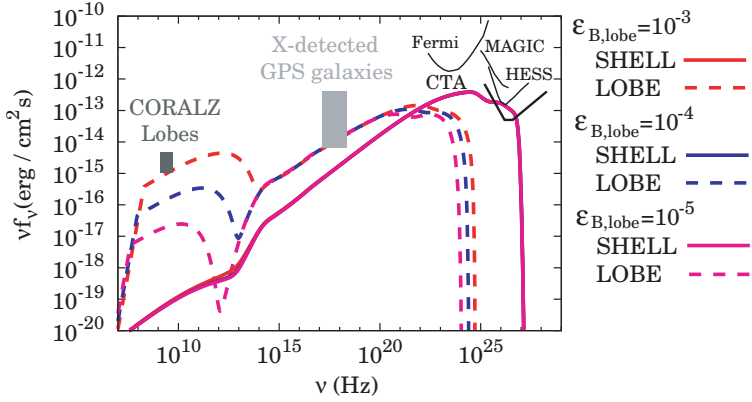}
\caption{
The same mini-shell and radio lobe spectra 
(solid and dashed curves, respectively)
but with
$L_{\rm UV}=2L_{\rm IR}=6\times 10^{44}~{\rm erg~s^{-1}}$, and
$\epsilon_{\rm B,lobe}=10^{-3},10^{-4},10^{-5}$.
The seed photons for IC scatttering in the mini-shell
are dominated by the core UV and IR photons.}
\label{fig:bright-core}
\end{figure}

In Fig.~\ref{fig:bright-core}, we further present
predicted spectra with larger
$L_{\rm UV}=2L_{\rm IR}=6\times 10^{44}~{\rm erg~s^{-1}}$, and
$\epsilon_{\rm B,lobe}=10^{-3}, 10^{-4}, 10^{-5}$.
Overall features of shell and lobe spectra are
similar to the cases in Figs. 1 and 2 
the mini-shell spectrum is detectable by CTA.
In this case, seed photons for IC scattering
are dominated by the emission of the nucleus 
$L_{\rm UV}=2L_{\rm IR}$.
Therefore, the shell IC luminosity in the TeV  band 
remains  constant for various $\epsilon_{\rm B, lobe}$.
Interestingly
the intrinsic absorption 
in the mini-shell produces the dip around 
$\sim 10^{25}~{\rm Hz}$.

\section{\bf Summary and discussions}

In the present work, we have studied 
non-thermal emissions from 
mini shells surrounding low-luminosity, 
mini radio lobes.
Predicted synchrotron emissions 
from mini-shells are very faint 
due to their weak magnetic field.
On the other hand,
IC scattering in the mini-shell
is significantly effective
since the energy densities of 
soft photons from AGN nucleus and
synchrotron photons from the radio lobe  
are large.
As a result, the IC emission from the shell 
overwhelms the emission 
from the radio lobe in VHE $\gamma$-ray range.
We find that
the non-thermal emission from the mini-shells
can be detectable by CTA and they
become a potential new class of VHE $\gamma$-ray emitters.

Among FIRST survey sources (White et al. 1997),
the candidate selection of CORALZs 
for further VLBI confirmation  had been done 
with the criteria of the flux density $>100$~mJy at 1.4~GHz
and the angular size $<2^{"}$ (de Vries 2009).
Hence a straightforward way to
increase the number of lower luminosity 
radio lobes is to conduct 
further  VLBI survey with a flux density threshold
lower than $100$~mJy among FIRST survey sources.
Then, lower luminosity lobes will be found more.
Another promising way to 
find less luminous mini radio lobes is a deep survey
with the next generation radio telescope SKA
with its collecting area distributed over
a large geographical area.
Such deep observations will significantly increase
the number of radio-dark mini-shells,
new class of VHE $\gamma$-ray emitters,
which can be potentially detected by CTA.


\bigskip
\leftline{\bf \large Acknowledgment}
\medskip

\noindent

We thank the referee for suggestions to 
improve the paper.
This work is partially supported by 
Grant-in-Aid for Scientific Research,
KAKENHI 24540240 (MK)
from Japan Society for the Promotion of Science (JSPS)
and 
by Research Activity Start-up 2284007 (NK) from the
Ministry of Education, Culture, Sports, Science, and Technology (MEXT).
Part of this work was done with the contribution
of the Italian Ministry of Foreign Affairs and Research
for the collaboration project between  Italy and Japan.



\footnotesize

\begin{thebibliography}{}

\bibitem[Abdo et al.(2009)]{2009ApJ...699...31A} Abdo, A.~A., Ackermann, 
M., Ajello, M., et al.\ 2009, \apj, 699, 31 


\bibitem[Actis et al.(2011)]{2011ExA....32..193A} Actis, M., Agnetta, G., 
Aharonian, F., et al.\ 2011, Experimental Astronomy, 32, 193 


\bibitem[Aharonian et al.(2006)]{2006Natur.440.1018A} Aharonian, F., 
Akhperjanian, A.~G., Bazer-Bachi, A.~R., et al.\ 2006, \nat, 440, 1018 




\bibitem[Aleksi{\'c} et 
al.(2012)]{2012A&A...539L...2A} Aleksi{\'c}, J., Alvarez, E.~A., Antonelli, L.~A., et al.\ 2012, \aap, 539, L2 




\bibitem[Bednarz 
\& Ostrowski(1998)]{1998PhRvL..80.3911B} Bednarz, J., \& Ostrowski, M.\ 1998, Physical Review Letters, 80, 3911 









\bibitem[Buttiglione et 
al.(2011)]{2011A&A...525A..28B} Buttiglione, S., Capetti, A., Celotti, A., et al.\ 2011, \aap, 525, A28 

\bibitem[Calderone et al.(2012)]{2012arXiv1204.2556C} 
Calderone, G., Sbarrato, T., \& Ghisellini, G. 2012, MNRAS, submitted
(arXiv:1204.2556)

\bibitem{CPD88} Carilli, C.~L., Perley, 
R.~A., \& Dreher, J.~H.\ 1988, \apjl, 334, L73 

\bibitem{CT02}Carilli, C. L., \& Taylor, G. B. 2002, ARA\&A, 40, 319

\bibitem{CMW75} Castor, J., McCray, R., 
\& Weaver, R.\ 1975, \apjl, 200, L107 


\bibitem[Coppi 
\& Aharonian(1997)]{1997ApJ...487L...9C} Coppi, P.~S., \& Aharonian, F.~A.\ 1997, \apjl, 487, L9 


\bibitem{CKH09} Croston, J.~H., et al.\ 2009, \mnras, 395, 1999 

\bibitem[Croston et al.(2007)]{2007ApJ...660..191C} Croston, J.~H., Kraft, 
R.~P., \& Hardcastle, M.~J.\ 2007, \apj, 660, 191 




\bibitem[de Vries et 
al.(2009)]{2009A&A...498..641D} de Vries, N., Snellen, I.~A.~G., Schilizzi, R.~T., Mack, K.-H., \& Kaiser, C.~R.\ 2009, \aap, 498, 641 





\bibitem{ESG01} Ellison, D.~C., Slane, P., \& Gaensler, B.~M.\ 2001, \apj, 563, 191 


\bibitem[Fanti et 
al.(1995)]{1995A&A...302..317F} Fanti, C., Fanti, R., Dallacasa, D., et al.\ 1995, \aap, 302, 317 


\bibitem[Fanti et 
al.(2011)]{2011A&A...528A.110F} Fanti, C., Fanti, R., Zanichelli, A., Dallacasa, D., \& Stanghellini, C.\ 2011, \aap, 528, A110 



\bibitem[Franceschini et 
al.(2008)]{2008A&A...487..837F} Franceschini, A., Rodighiero, G., \& Vaccari, M.\ 2008, \aap, 487, 837 






\bibitem[Funk 
\& Hinton(2012)]{2012arXiv1205.0832F} Funk, S., \& Hinton, J.\ 2012, 
Astroparticle Phys. in press (arXiv:1205.0832)





\bibitem[Guainazzi et 
al.(2006)]{2006A&A...446...87G} Guainazzi, M., Siemiginowska, A., Stanghellini, C., et al.\ 2006, \aap, 446, 87 







\bibitem[Holt et al.(2007)]{2007MNRAS.381..611H} Holt, J., Tadhunter, 
C.~N., Gonz{\'a}lez Delgado, R.~M., et al.\ 2007, \mnras, 381, 611 


\bibitem[Holt et al.(2009)]{2009MNRAS.400..589H} Holt, J., Tadhunter, 
C.~N., \& Morganti, R.\ 2009, \mnras, 400, 589 





\bibitem[Ito et al.(2011)]{2011ApJ...730..120I} Ito, H., Kino, M., 
Kawakatu, N., \& Yamada, S.\ 2011, \apj, 730, 120 (I11)






\bibitem[Kawakatu et al.(2009)]{2009ApJ...693.1686K} Kawakatu, N., Nagao, 
T., \& Woo, J.-H.\ 2009, \apj, 693, 1686 

\bibitem[Kino et al.(2009)]{2009MNRAS.395L..43K} Kino, M., Ito, H., 
Kawakatu, N., \& Nagai, H.\ 2009, \mnras, 395, L43 


\bibitem[Kino et al.(2007)]{2007MNRAS.376.1630K} Kino, M., Kawakatu, N., 
\& Ito, H.\ 2007, \mnras, 376, 1630 




\bibitem[Koyama et al.(1995)]{1995Natur.378..255K} Koyama, K., Petre, R., 
Gotthelf, E.~V., et al.\ 1995, \nat, 378, 255 

\bibitem[Kunert-Bajraszewska 
\& Labiano(2010)]{2010MNRAS.408.2279K} Kunert-Bajraszewska, M., \& Labiano, A.\ 2010, \mnras, 408, 2279 




\bibitem[Lazio(2011)]{} Lazio, J. on behalf of Science Working Group,
The Square Kilometre Array Design Reference Mission: SKA Phase I,
(SCI-020.010.020-DRM-002)



\bibitem[Mack et al.(2009)]{2009AN....330..217M} Mack, K.-H., Snellen, 
I.~A.~G., Schilizzi, R.~T., 
\& de Vries, N.\ 2009, Astronomische Nachrichten, 330, 217 


\bibitem{MB03} Mathews, W.~G., \& Brighenti, F.\ 2003, \araa, 41, 191 

\bibitem[McConville et al.(2011)]{2011ApJ...738..148M} McConville, W., 
Ostorero, L., Moderski, R., et al.\ 2011, \apj, 738, 148 



\bibitem[Nagai et al.(2010)]{2010PASJ...62L..11N} Nagai, H., Suzuki, K., 
Asada, K., et al.\ 2010, \pasj, 62, L11 


\bibitem{OMS10} Ostorero, L., et al.\ 2010, \apj, 715, 1071 

\bibitem{OM88} Ostriker, J.~P., \& McKee, C.~F.\ 1988, 
Reviews of Modern Physics, 60, 1 

\bibitem[Orienti et al.(2010)]{2010MNRAS.408.1075O} Orienti, M., Dallacasa, 
D., \& Stanghellini, C.\ 2010, \mnras, 408, 1075 













\bibitem[Snellen et al.(2004)]{2004MNRAS.348..227S} Snellen, I.~A.~G., 
Mack, K.-H., Schilizzi, R.~T., \& Tschager, W.\ 2004, \mnras, 348, 227 

\bibitem[Snellen et al.(1999)]{1999MNRAS.307..149S} Snellen, I.~A.~G., 
Schilizzi, R.~T., Bremer, M.~N., et al.\ 1999, \mnras, 307, 149 


\bibitem[Son et al.(2012)]{2012ApJ...757..140S} Son, D., Woo, J.-H., Kim, 
S.~C., et al.\ 2012, \apj, 757, 140 







\bibitem{SBM08} Stawarz, {\L}., 
Ostorero, L., Begelman, M.~C., Moderski, R., Kataoka, J., 
\& Wagner, S.\ 2008, \apj, 680, 911 

\bibitem[Su et al.(2010)]{2010ApJ...724.1044S} Su, M., Slatyer, T.~R., 
\& Finkbeiner, D.~P.\ 2010, \apj, 724, 1044 


\bibitem[Suzuki et al.(2012)]{2012ApJ...746..140S} Suzuki, K., Nagai, H., 
Kino, M., et al.\ 2012, \apj, 746, 140 



\bibitem[Tengstrand et 
al.(2009)]{2009A&A...501...89T} Tengstrand, O., Guainazzi, M., Siemiginowska, A., et al.\ 2009, \aap, 501, 89 


\bibitem[Vink et al.(2006)]{2006MNRAS.367..928V} Vink, J., Snellen, I., 
Mack, K.-H., \& Schilizzi, R.\ 2006, \mnras, 367, 928 







\end{thebibliography}




\begin{table}
\centering
\caption{Parameters of Shell and Radio lobes}
\label{tab:parameters}       
\begin{tabular}{lcl}
\hline\noalign{\smallskip}
{\bf parameters} & {\bf symbols} & {\bf values}\\
\noalign{\smallskip}\hline\noalign{\smallskip}

 jet power&
 $L_{\rm j}$ &
 $ 5\times 10^{46}~{\rm ergs~s^{-1}}$
 \\
 Distance from core to  shell &
 $R $ &
 $5~{\rm pc}$  
 \\
 luminosity of IR emissions from dust-torus &
 $L_{\rm IR}$ &
$ 3 \times (10^{44}, 10^{43}, 10^{42})~{\rm ergs~s^{-1}}$ 
 \\
 luminosity of UV emissions from accretion  disk &
 $L_{\rm UV}$ & 
$6\times (10^{44}, 10^{43}, 10^{42})~{\rm ergs~s^{-1}}$ 
\\
fraction of non-thermal electrons &
 $\epsilon_{\rm e,shell}$ &
 0.05
 \\
power-law index of injected electrons (shell)&
 $p_{\rm shell}$ &
 $2$
 \\
gyro-factor &
 $\xi_{\rm shell}$ &
 10
\\
 redshift &
 z &
 0.08  
\\
 B energy fraction&
 $\epsilon_{\rm B,lobe}$ &
 $10^{-3}$,  $10^{-4}$,  $10^{-5}$ 
 \\
 fraction of non-thermal electrons&
 $\epsilon_{\rm e,lobe}$ &
 $10^{-2}$
 \\
 power-law index of injected electrons&
 $p_{\rm lobe}$ &
 $2.2$
 \\
gyro-factor &
 $\xi_{\rm lobe}$ &
 $10^{7}$ 
\\ 
\noalign{\smallskip}\hline
\end{tabular}
\end{table}

\end{document}